# SPECTRAL CLASSIFICATION OF GALAXIES

Laerte Sodré Jr. & Héctor Cuevas
Departamento de Astronomia
Instituto Astronômico e Geofísico da Universidade de São Paulo
Av. Miguel Stefano 4200, 04301-904 São Paulo, Brazil
e-mail: laerte@astro1.iagusp.usp.br hcuevas@astro1.iagusp.usp.br

**Abstract** – We investigate the integrated spectra of a sample of 24 normal galaxies. A principal component analysis suggests that most of the variance present in the spectra is due to the differences in morphology of the galaxies in the sample. We show that spectroscopic parameters extracted from the spectra, like the amplitude of the 4000 Å break or of the CN band, correlate well with Hubble types and are useful for quantitative classification.

**key-words** – galaxies – spectral classification – Hubble sequence

## 1. INTRODUCTION

The integrated spectrum of a galaxy represents a weighted mean in luminosity of the stellar populations that make it up. The integrated spectra of nearby galaxies show remarkable regularity, correlating very well with the Hubble types: galaxies of same morphological type tend to have similar stellar populations. It is this regularity that allows one to classify galaxies by its spectral type, as done since the pioneering works of Humason (1936) and Morgan & Mayall (1957).

Spectral classification presents an important advantage when compared to morphological classification: it can be estimated up to much larger distances. It is almost impossible to ascribe reliable morphological types to normal galaxies farther than $z \sim 0.5$ due to the combined effect of seeing and $(1+z)^4$ cosmological surface brightness dimming. Therefore, spectral types may be an attractive alternative to Hubble types for studies in observational cosmology.

The study of the distribution of spectral types for large samples of field and cluster galaxies, for instance, may help to explain the nature of the density-morphology relation (Dressler 1980) or to determine the dependence of the luminosity of galaxies with morphology and environment.

The spectral type may be estimated objectively from the integrated spectrum of a galaxy and characterizes its photometrically dominant stellar populations. Bica (1988) proposes a classification system based on spectra of the central regions of galaxies. For a recent application of spectral classification to high redshift galaxies, see Dressler & Gunn (1992). Here we investigate integrated spectra of normal galaxies at low redshift, looking for features useful for spectral classification.

## 2. SPECTRAL PARAMETERS

The hypothesis that underlies spectral classification is that galaxies actually form a kind of one-dimensional sequence in the data space representing their spectra, and that this sequence runs along the Hubble morphological sequence. This data space may be such that an entire spectrum is represented by a single point. However, it is both possible and desirable to deal with spaces of much lower dimensionality, by representing a spectrum with fewer parameters. This representation may be very crude (*e.g.*, by using colours), or more detailed, if the spectra are described by a set of suitably chosen spectroscopic index. A critical stage in the design of a classifier is *feature selection*, that is, the choice of quantities or qualities that are most suited for the description of the morphology of galaxies. Useful features should be easy to measure and must be able to provide accurate classification. We will call *spectral parameters* the features extracted from the spectra and used in spectral classification.

We are currently investigating the quantitative estimation of spectral types by using data from Kennicutt (1992). He measured integrated spectra of 55 normal and peculiar galaxies, with 5-8 Å resolution. Here we discuss some characteristics of the integrated spectra of 24 normal galaxies. A study of the spectrophotometric properties of all the 55 galaxies is in progress and will be published elsewhere. This sample of normal galaxies covers the Hubble sequence from E to Im, avoiding any evidence of peculiarity (*e.g.*, AGNs, starbursts, mergers). Although small, this sample allows one to explore the dependence of the spectrophotometric properties of a galaxy with its morphology.

First we present results of a Principal Component Analysis (PCA) of the spectra. PCA is a useful technique for reducing the dimensionality of a data space by identifying the linear combinations of input parameters with maximum variance, the 'principal components' (Murtagh & Heck 1987). Our anal-ysis was done on the correlation matrix of the spectra with 1300 channels over the rest-frame wavelength interval 3750-6500Å. For these normal galaxies the first 5 principal components accounts for 90% of the sample variance. It is reasonable to assume that most of the variance is due to the morphological mix of the sample. Indeed, Spearman's rank-order correlation coefficient between Hubble types and the first principal component is high, 0.87, indicating that this correlation is significant (figure 1). For a similar application of PCA to QSO spectra, see Francis *et al.* (1992).

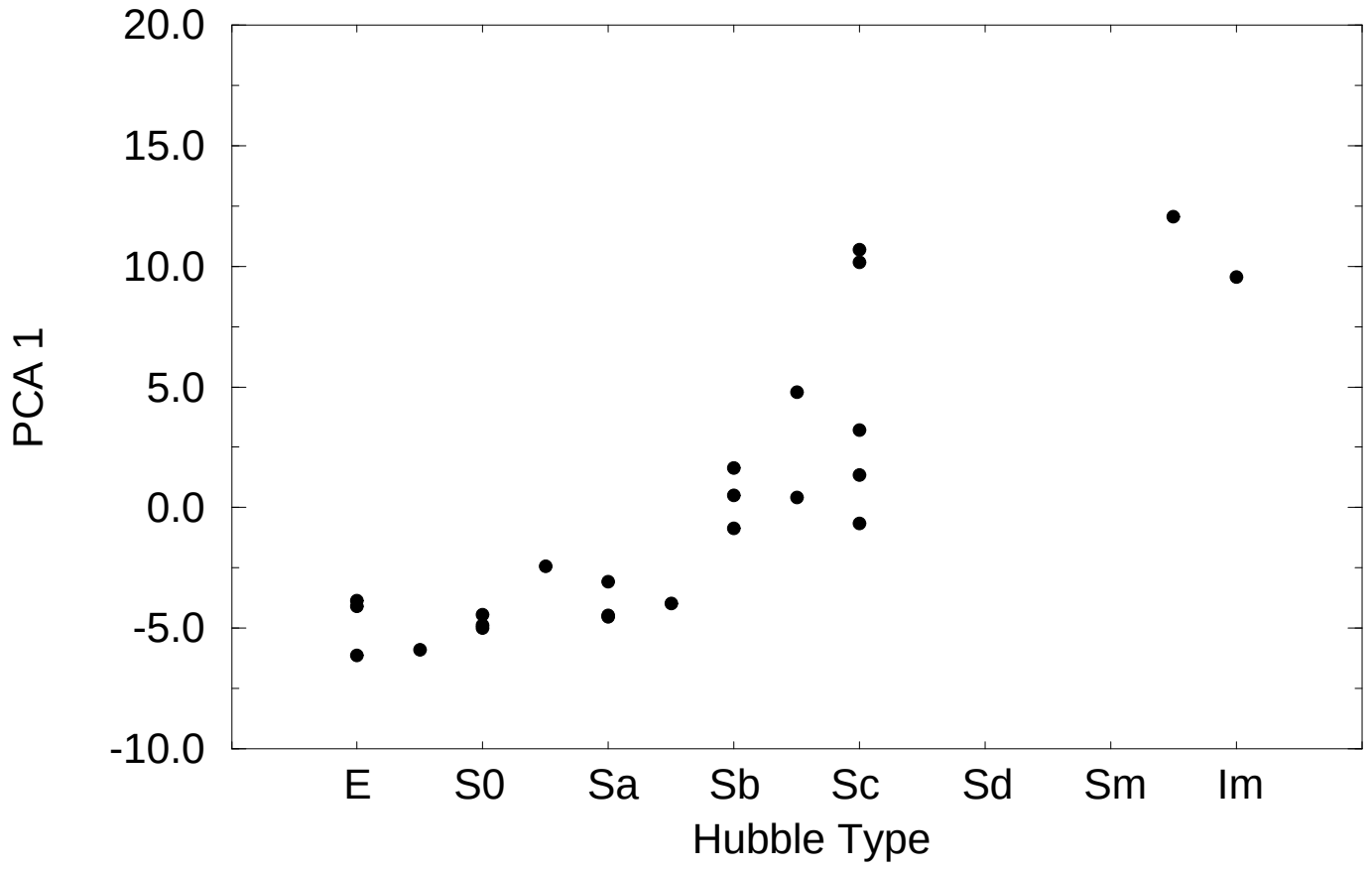

Figure 1: Projection of 1300 channels (3750Å to 6500Å) of the normal galaxy spectra on the first principal component, versus the Hubble type.

The information carried by a spectrum comes from the lines as well as from the continuum. The differences in the continuum of normal galaxy spectra imply that colours may provide reliable spectral classification. Broad band colours like $(U - B)$ or $(B - V)$ have been recognized as good morphological indicators since the classical study done by Holmberg (1958) and a set of colours may be considered as a very low resolution spectrum.

An interesting class of parameters are the spectroscopic indices, like the amplitude $\Delta$ of the 4000Å break and the strength of the G band or the Mg2 index. We follow here the definitions of Brodie & Huchra (1990). These indices are metalicity or temperature indicators and are useful in the diagnosis of stellar populations in complex stellar systems. Figure 2 shows that these indices correlate very well with morphological types. The break amplitude $\Delta$, for instance, decreases from $\sim$2, for the ellipticals, to $\sim$1, for Sm/Im galaxies.

## 3. DISCUSSION

By spectral classification we mean to attribute a Hubble type to a galaxy from parameters extracted from its integrated spectrum. It is possible to classify galaxies this way using either the continuum, as measured by colours or very low resolution spectra, or line indices, for spectra with better resolution.

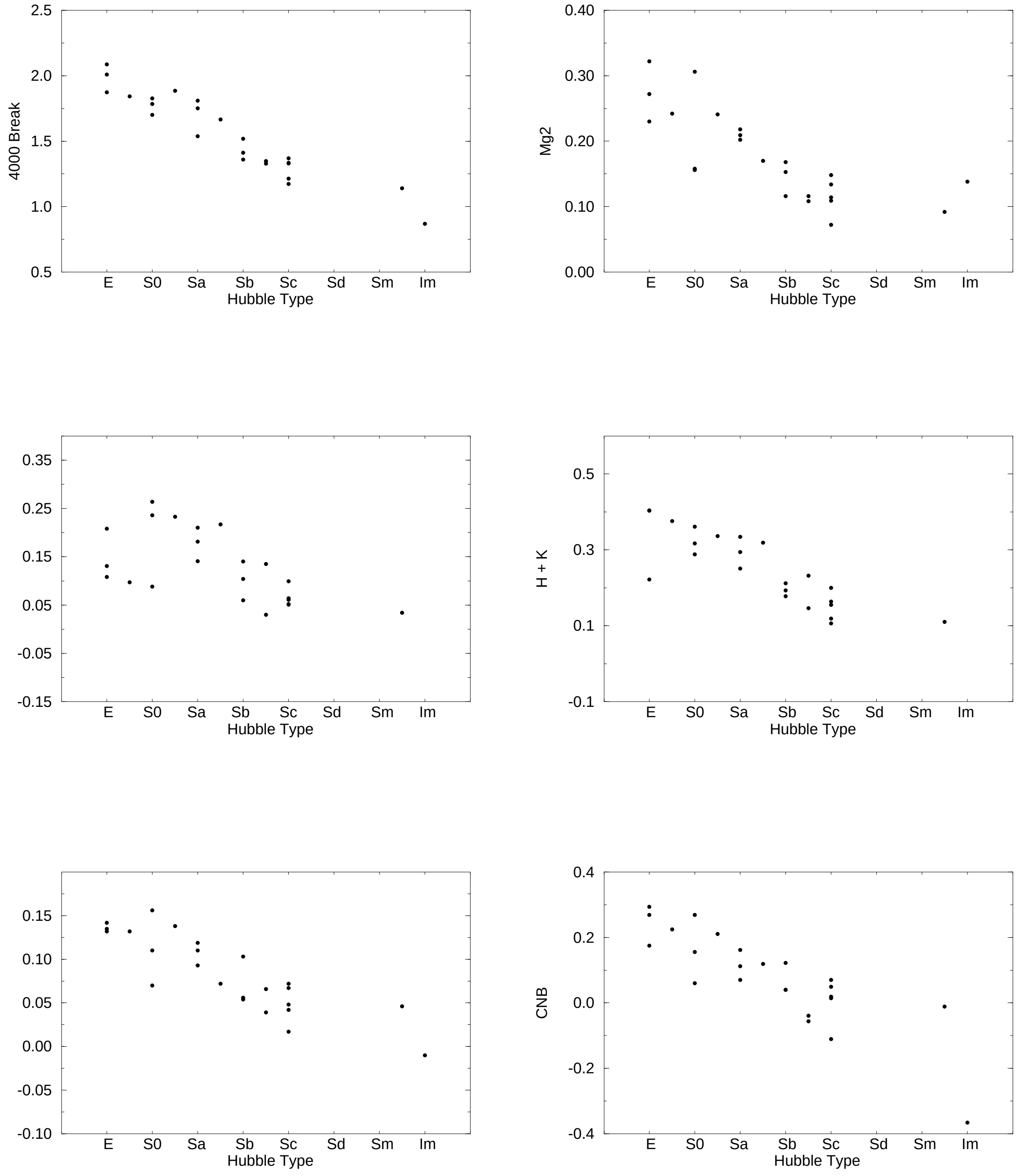

Figure 2: Some spectral parameters (see Brodie & Huchra 1990) versus Hubble type for the normal galaxy sample.

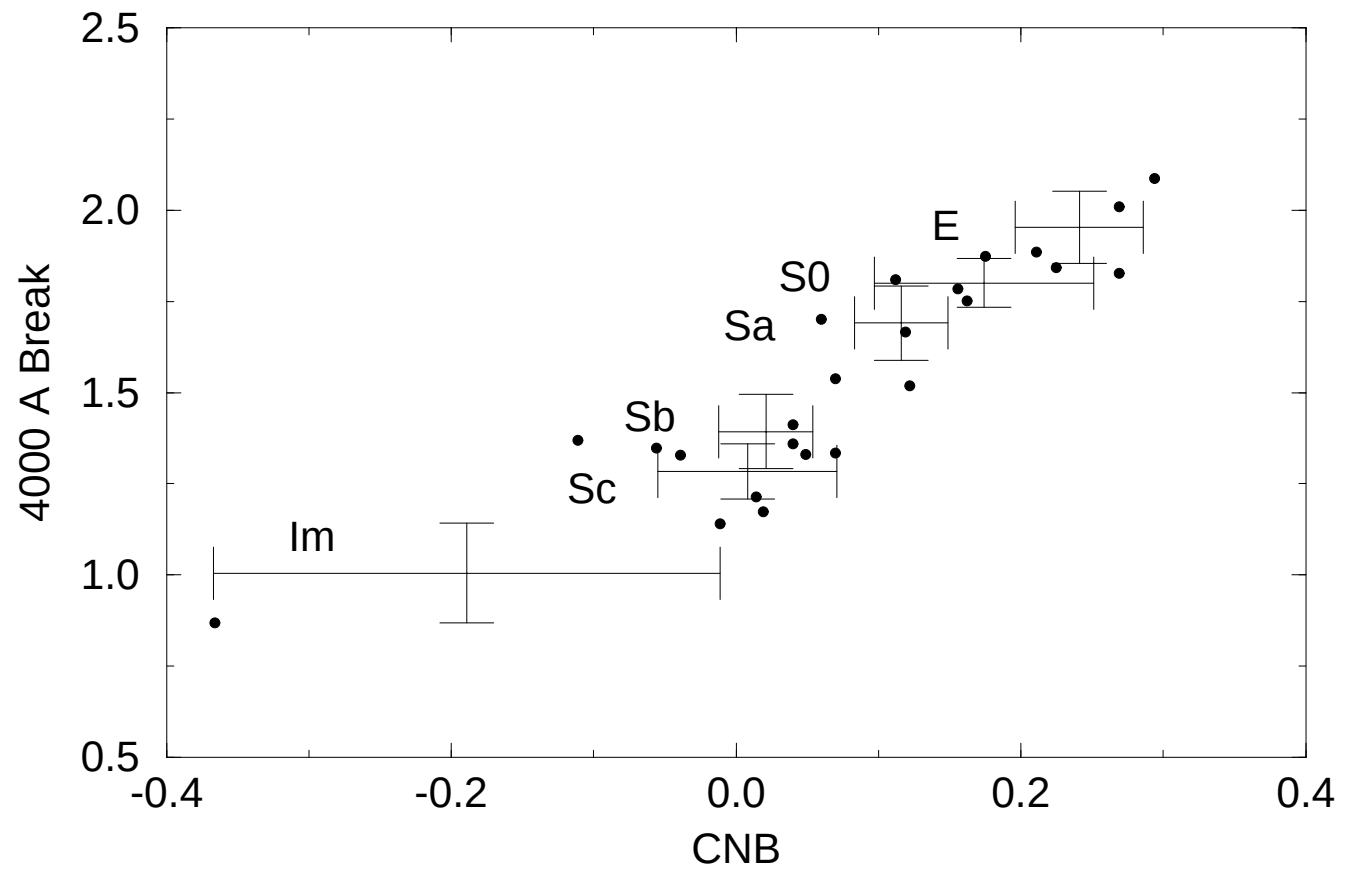

Figure 3: Amplitude of the 4000Å break versus the CNB index (dots). Mean values for each morphological group with the 1 $\sigma$ standard deviation for each parameter are also shown.

There is a clear monotonic variation between the spectral parameters and morphological type, as shown in figure 2. As these parameters are measures of stellar populations, they are expected to correlate with the history of star formation in galaxies, which underlies the Hubble sequence. We plot in figure 3 an index that measures the intensity of the CN band (the CNB index) against the break amplitude, indicating also the mean locus of each morphological type. Figure 3 shows that, at least for the small sample of normal galaxies analysed here, the morphological sequence corresponds to a one-dimensional sequence in data space, giving then 'topological' support to the Hubble sequence.

Spectral classification may prove useful in the analysis of spectra taken for new generation redshift surveys, like the 2dF UKGG survey, which aims to measure the spectra of 250,000 galaxies (Lahav 1994). Another valuable material for spectral classification are galaxy surveys with objective prism. Note that a slitless spectrum provides, if properly extracted, a better approximation to an integrated spectrum than ordinary slit spectroscopy, which probes essentially the central regions of galaxies. These surveys can be used to study galaxy morphology up to moderate redshifts.

Spectral parameters of galaxies of known morphology can be used to train an artificial neural network (ANN), which can then be used to give types to large amounts of spectra of galaxies without previous classification. ANNs are interesting for classification due to their ability to learn, to generalize, and to cluster or organize data (Hertz *et al.* 1991). Recent applications of ANNs in Astronomy that are relevant for the present work are morphological classifi-cation of galaxies (Storrie-Lombardi *et al.* 1992) and spectral classification of stars (von Hippel *et al.* 1994). Automated procedures like ANNs are the only practical way of classifying the enormous amount of data already available or to be produced by the surveys now in preparation.

Due to their nature, spectral indices are more prone to be related to the physics underlying the Hubble sequence than the principal components. But the use of principal components as inputs for classification with ANNs shows some advantages. For instance, they allow to avoid the 'curse of dimensionality', since instead of feeding a neural net with, say, 1300 values of spectroscopic channels, one may use a dozen or so inputs carrying essentially the same amount of information. The number of weights in the ANN in the latter approach is much smaller than in the former, making classifications more robust and stable. Additionally, by using principal components we allow the data to speak by itself, without the need of pre-judging whether, say, a given spectral index is better than another.

## 4. CONCLUSION

We used the integrated spectra of a sample of normal galaxies to investigate the relation of spectral parameters with Hubble types. Our results show that several spectrophotometric indices correlate well with morphological types and, consequently, they are useful for galaxy classification.

The spectral classification of large magnitude or diameter selected samples may provide interesting results for observational cosmology. For instance, we may be able to study correlations between spectroscopic parameters and luminosity for galaxies of all morphological types in different environments, and investigate the presence of intermediate age stellar populations in early type systems. Furthermore, the accurate knowledge of the distribution of spectroscopic parameters for local galaxies may prove essential to quantify the amount of spectrophotometric evolution and the nature of high redshift galaxy populations.

**Acknowledgements** – We would like to thank the financial support provided by FAPESP and CNPq. LSJ thanks the hospitality of the Institute of Astronomy and the Royal Greenwich Observatory, Cambridge, England, where part of this work was done.